\documentclass[letterpaper,notitlepage,11pt]{article}
\usepackage{amsmath,amssymb}
\usepackage{slashed}
\usepackage[utf8]{inputenc}
\usepackage{hyperref}
\usepackage{cite} 
\usepackage[margin=1in]{geometry}
\usepackage[usenames]{color}
\usepackage{ulem}
\usepackage{authblk}

\def\bea{\begin{eqnarray}}
\def\eea{\end{eqnarray}}
\def\bean{\begin{equation*}}
\def\eean{\end{equation*}} 
\def\beq{\begin{equation}}
\def\eeq{\end{equation}} 

\begin{document}
  \title{Comment on \\ ``Neutron lifetime and dark decay of the neutron and hydrogen''}
\author{Bartosz Fornal and Benjam\'{i}n Grinstein}
\affil{\small {\textit{Department of Physics, University of California, San Diego, \\ \textit{9500 Gilman Drive, La Jolla, CA 92093, USA}}}}
\date{\small\today}
\maketitle

\begin{abstract}
\vspace{2mm}
The manuscript by Berezhiani (arXiv:1812.11089) proposes a model that has the neutron decaying into a mirror neutron
with a branching fraction of 1\%, alleviating the tension between neutron lifetime measurements in
beam {\it vs} bottle experiments.  We show that the model as proposed is inconsistent with experiment. Variations of the model may work at the expense of extreme breaking of the $Z_2$ symmetry between the Standard Model and its mirror copy.\\
\end{abstract}

Slightly over a year ago we proposed that the tension between neutron lifetime
measurements in beam {\it vs} bottle experiments can be alleviated by an unobserved decay of the neutron into new neutral particles with a branching fraction of 1\% \cite{Fornal:2018eol}. To  describe one of the proposed putative new decay channels ($n\rightarrow \chi\,\gamma$) in a quantitative way, we considered theories in which a dark Dirac fermion $\chi$ has a small mass mixing with the neutron. Diagonalizing the mass matrix gives rise to an interaction $ \chi\, n\, \gamma$. An example of such a theory is given by the effective Lagrangian
\beq\label{F0}
\mathcal{L}^{\rm eff}= \bar{n}\left(i\slashed\partial-m_n +\frac{g_ne}{2 m_n}\sigma^{\mu\nu}F_{\mu\nu}\right) n 
+ \   \bar{\chi}\left(i\slashed\partial-m_\chi\right) \chi + \varepsilon \left(\bar{n}\chi + \bar{\chi}n\right) \ ,
\eeq
where $\varepsilon$ is the mass-mixing parameter. Transforming to the mass eigenstate basis yields,  for \linebreak $\varepsilon \ll m_n-m_\chi$\,,
\bea\label{eff1c0}
\mathcal{L}_{n \rightarrow \chi \gamma}^{\rm eff} =\frac{g_n e}{2m_n}\frac{\varepsilon}{(m_n-m_\chi)}\,\bar\chi \,\sigma^{\mu\nu} F_{\mu\nu} \,n \ .
\eea
Therefore, the neutron dark decay rate is
\bea\label{rate-phase000}
\Delta\Gamma_{n\to\chi\gamma} = \frac{g_n^2e^2}{8\pi}\bigg(1-\frac{m_\chi^2}{m_n^2}\bigg)^{\!3}  \frac{m_n\,\varepsilon^2}{(m_n-m_\chi)^2}
\, \approx \,   \Delta \Gamma_{n}^{\rm exp} \,\left(\frac{1+x}{2}\right)^{\!3}\! \left(\frac{1-x}{1.8\times 10^{-3}}\right)\!\left(\frac{\varepsilon \ [{\rm GeV}]}{9.3\times 10^{-14}}\right)^{\!2},
\eea
where $x=m_\chi/m_n$. Stability of $^9\text{Be}$ with respect to the potential  decay channel $^9\text{Be} \to \,^8\text{Be}+\chi$ requires
$m_\chi>937.900\;\text{MeV}$. A slightly stronger constraint follows from
$^9\text{Be}\;\slashed{\to}\;2\,^4\text{He}+\chi$ \cite{Pfutzner:2018ieu}:
  \bea\label{constr}
937.992 \ {\rm MeV} <  m_\chi <  939.565 \ {\rm MeV} \ . 
\eea

A minimal particle physics realization of this idea was also presented in
Ref.~\cite{Fornal:2018eol}. It requires only two particles
beyond the Standard Model (SM): a complex scalar $S\sim(3,1)_{-1/3}$, and a Dirac fermion $\chi\sim(1,1)_0$, with the Lagrangian
\bea\label{Eq:Lag1}
\mathcal{L}_{1} =  \big(\lambda_q \,\epsilon^{ijk}\, \overline{u^c_L}_{i}\, d_{Rj}  S_k + \lambda_\chi S^{*i}\bar\chi \,d_{Ri} + {\rm h.c.}\big) - M_ S^2 | S|^2 - m_\chi \,\bar\chi\,\chi  \ , 
\eea
where $u^c_L$ is the charge conjugate of $u_R$. With $B_\chi\!=\!1$ and $B_
S=-2/3$, baryon number  is conserved. 

\pagebreak
\noindent
The rate for $n \!\rightarrow \!\chi \,\gamma$ is given by Eq.\,(\ref{rate-phase000}) with
\bea\label{Eq:our-epsilon}
\varepsilon = \frac{\beta\,\lambda_q\lambda_\chi }{M_{ S}^2}\,.
\eea
Here $\beta$ is defined by
\(\langle 0| \epsilon^{ijk} \!\left(\overline{u^c_L}_{i} d_{Rj}\right)
\!d_{Rk} |n\rangle = \tfrac12\beta \left(1+\gamma_5\right) u_n \)\,, where 
$u_n$ is the neutron Dirac spinor. Lattice QCD calculations give
$\beta = 0.0144(3)(21) \ {\rm GeV}^3$ \cite{Aoki:2017puj}.  Assuming
$m_\chi\approx 938 \ {\rm MeV}$ to maximize the rate, the parameter choice
explaining the anomaly is
\(
|\lambda_q\lambda_\chi|/M_ S^2  \approx 7 \times 10^{-6}  \  {\rm TeV}^{-2} 
\).
\\

After a long introduction that reviews the work above\footnote{Without proper accreditation to
  Refs.~\cite{Fornal:2018eol} and \cite{Pfutzner:2018ieu}.}, Ref.~\cite{Berezhiani:2018udo} proposes
a model in which $\chi$ is identified as the ``neutron'' ($n'$) of a mirror copy of the SM. The
complete model is a double replica of the model in Eq.\,\eqref{Eq:Lag1},
extending it by mirroring all SM
particles and the new particle $S$, and duplicating  the
Lagrangian, with $\chi$ as a mediator:
\bea\label{Eq:lag2}
\mathcal{L}_{2} = \lambda_q
\,\epsilon^{ijk}\, \overline{u^c_L}_{i}\, d^{\phantom{c}}_{Rj}  S_k + \lambda_\chi
S^{*i}\bar\chi \,d^{\phantom{c}}_{Ri}
+\lambda_q \,\epsilon^{ijk}\, \overline{u^{\prime c}_L}_{i}\, d'_{Rj}  S'_k + \lambda_\chi S^{\prime*i}\overline{\chi^c} \,d'_{Ri} + {\rm h.c.}\ .
\eea
Integrating out $S$, $S'$
and $\chi$ produces the effective operator for the $n-n'$ mixing with
\bea\label{Eq:his-epsilon}
\varepsilon = \frac{(\beta\,\lambda_q\lambda_\chi)^2}{M_{ S}^2 M_{ S'}^2m_\chi}\,.
\eea
This is to be compared with Eq.\,\eqref{Eq:our-epsilon}. If the $Z_2$ symmetry
is explicit,  i.e.~$M_{S'}=M_S$, then in order to have
$\varepsilon\approx10^{-13}~\text{GeV}$, one requires
$M_{S,S'} \approx (210~\text{GeV}) \sqrt{\lambda_q\lambda_\chi} /(m_\chi[\text{GeV}])^{1/4}$. To
avoid light colored scalars, $Z_2$ must be spontaneously broken. Following
Ref.~\cite{Berezhiani:2018udo}, i.e.~taking $M_S/\lambda_q\gtrsim10$~TeV and $m_\chi=10$~GeV, 
Eq.\,\eqref{Eq:his-epsilon} results in\footnote{Our value differs from that of
  Ref.~\cite{Berezhiani:2018udo} by a factor of 20, of which a factor
  of 10 results from the overestimate of $\beta$ used there.}  $M_{S'}/\lambda_\chi
\lesssim1.4$~GeV.
More conservatively, it is sufficient to have  $m_\chi\gtrsim1$~GeV. Using
the bound from LHC dijet searches for scalar diquarks, $M_S/\lambda_q \gtrsim 7.0~\text{TeV}$
\cite{Sirunyan:2018xlo,Aaboud:2017yvp}, results in
$M_{S'}/\lambda_\chi\lesssim6.5$~GeV. A complimentary bound on $M_S$ alone comes from investigating the four-jet signature of scalar diquarks \cite{Richardson:2011df,Assad:2017iib} and gives $M_S \gtrsim 2.5~\text{TeV}$.

Even if spontaneously broken, the $Z_2$ symmetry insures that the UV
value of the gauge and Yukawa coupling constants in the SM and its
mirror copy are the same. The low-energy values of the gauge
couplings are determined by the renormalization group with the
equality of couplings as a UV boundary condition,
i.e.~$g'_3(\mu)=g_3(\mu)$ for $\mu \gg M_S$, assuming $S$ is the
heaviest field. We then have at one loop (sufficiently accurate for
the point we intend to make): \bea\label{Eq:ratio} \frac{\Lambda_{\rm
    QCD'}^{(3)}}{\Lambda_{\rm
    QCD}^{(3)}}=\left(\frac{M_{S'}}{M_S}\right)^{\frac1{54}} .  \eea
Here
$\alpha_s(\mu)=2\pi/\big[\beta_0\ln(\mu/\Lambda_{\rm
    QCD}^{(n_f)})\big]$, with $\Lambda_{\rm QCD}^{(n_f)}$ being the
one-loop RG-invariant scale for $n_f$ quarks, and similarly for the
mirror sector. The ratio in Eq.\,\eqref{Eq:ratio} is independent of
the subtraction scheme at one loop. For the most conservative choice,
i.e.~$M_S \approx2.5 \ {\rm TeV}$,
  $M_S/\lambda_q \approx7.0 \ {\rm TeV}$ and therefore
  $M_{S'}\approx(6.5 \ {\rm GeV})\lambda_\chi\le
  61~\text{GeV}$ (for $\lambda_\chi\le 3\pi$), then this gives
$\Lambda_{\rm QCD'}^{(3)}/\Lambda_{\rm
    QCD}^{(3)}\lesssim0.934$, which immediately translates into
$m_{n'}\lesssim 0.934\,m_n=878~\text{MeV}$.
If, following Ref.~\cite{Berezhiani:2018udo}, one instead takes
$m_\chi \approx 10$~GeV and
$M_S/\lambda_q\approx10 \ {\rm TeV}$, which requires (using the
correct value for $\beta$)
$M_{S'}\lesssim (1.5~\text{GeV}) \lambda_\chi<14~\text{GeV}$ (for
$\lambda_\chi\le 3\pi$), then, given that
  $M_S \gtrsim 2.5 \ {\rm TeV}$, the mirror neutron mass is \bea
m_{n'}\lesssim 854~\text{MeV}\ .  \eea These results
violate the bound in Eq.\,\eqref{constr}, giving an unstable
$^9\text{Be}$ and, in addition, an unstable proton. Therefore, the
model as presented in Ref.~\cite{Berezhiani:2018udo} is inconsistent
with observation.\pagebreak

A possible way to avoid this disastrous conclusion is to use the
breaking of the $Z_2$ symmetry to invoke large $u'$, $d^{\:'}$ and
$s'$ masses, that could then increase $m_{n'}$ to within
$\sim1~\text{MeV}$ of $m_n$. Using \bea \sigma_{\pi N}\equiv
\tfrac12(m_u+m_d)\langle N|(\bar uu+\bar d
d)|N\rangle\qquad\text{and}\qquad \sigma_{s N}\equiv m_s\langle N|\bar
ss|N\rangle\ , \eea we estimate the additional quark mass contribution
to the mirror neutron mass as
\bea \Delta m_{n'} &\approx&
\left(\frac{m_{u'}+m_{d'}}{m_u+m_d}-1\right)\sigma_{\pi N}+
  \left(\frac{m_{s'}}{m_s}-1\right)\sigma_{s N} \ ,
\eea
where $\sigma_{\pi N} \approx 46~\text{MeV}$ and  $\sigma_{s N} \approx 40~\text{MeV}$ from the lattice
calculation in Ref.~\cite{Yang:2015uis}. Similar values for
$\sigma_{\pi N}$ and $\sigma_{sN}$ are found elsewhere
\cite{Durr:2011mp,Horsley:2011wr,Alexandrou:2014sha,Bali:2012qs,Durr:2015dna};
the values we use are at the upper end of the range in these
determinations, so that our estimate of mirror quark masses required
to raise the mirror neutron mass is conservatively low.\footnote{In
  addition, the linear relation $m_N=m_0+a m_q$ is only a rough
  approximation. Corrections of order $m_\pi^3\propto m_q^{3/2}$ come
  in with a negative sign and are not negligible.}  The $Z_2$
symmetry requires equality of Yukawa couplings that yield quark and
mirror quark masses; hence, the mirror quark to quark mass ratios are simply given by
the ratio of electroweak vacuum expectation values in the mirror SM to
that of the SM, $v'/v$. The masses of heavy mirror quarks also change
and the estimate of $\Lambda'$ must be revisited:
\bea\label{Eq:ratio2}
\frac{\Lambda_{\rm QCD'}^{(3)}}{\Lambda_{\rm QCD}^{(3)}}=
\left(\frac{v'}{v}\right)^{\frac{2}{9}}
\left(\frac{M_{S'}}{M_S}\right)^{\frac1{54}}.
\eea
In addition, we must
distinguish $\beta'$ from $\beta$ in Eq.\,\eqref{Eq:his-epsilon}, with
$\beta'\approx
(\Lambda_{\rm QCD'}^{(3)}/\Lambda_{\rm QCD}^{(3)})^3\beta$. For the values
$\varepsilon=10^{-13}~\text{GeV}$, $M_S=2.5~\text{TeV}$, $M_S/\lambda_q = 7.0 \ {\rm TeV}$, $m_\chi=1.0~\text{GeV}$ and
$\lambda_\chi=1$, and setting $m_{n'}=938.8~\text{MeV}$ (i.e. at the lower bound arising  from the search for radiative $n$ dark decay
\cite{Tang:2018eln}), we find $v'/v=1.4$ and
$M_{S'}=6.1~\text{GeV}$.

A mirror world with $v'/v\sim2$ is likely to be very different from ours. It has been argued that
mirror deuteron is unstable to weak decay for $v'/v\gtrsim1.2$ and unstable to strong decay for
$v'/v\gtrsim1.4$; these estimates are based on a square well model; when the OBEP model is used
instead, the threshold values of $v'/v$ for the weak and strong decays  become 1.4 and 2.7,
respectively \cite{Agrawal:1998xa,Agrawal:1997gf}. Because our estimates for $v'/v$ are in this range, we cannot draw solid
conclusions;  while it was not pressing for the work in Refs.~\cite{Agrawal:1998xa,Agrawal:1997gf}, a refined
calculation that can sharply establish the threshold for mirror deuterium stability is needed in our
context to determine the fate of mirror deuterium and other mirror nuclei (and the cosmological
implications of the mirror world). \vspace{4mm}

We have explored but one consequence of a badly broken $Z_2$ symmetry in the model of
Ref.~\cite{Berezhiani:2018udo}.  There may be other effects that as yet have not been accounted
for. But given that $Z_2$ is so badly broken, rather than mapping out the consequences of the
broken symmetry, it may be more productive to reinterpret our result
in a more general context: instead of restricting the dark sector to a $Z_2$ symmetric mirror SM,
one may more generally conceive of models with a dark sector involving composite states,  bound by a
dark-strong interaction, not necessarily based on an $SU(3)$ gauge group nor on ``quarks'' in the
fundamental representation. For example,
one may replace the mirror sector in the Lagrangian in Eq.\,\eqref{Eq:lag2} by a simpler model:
\bea\label{Eq:lag3}
\mathcal{L}_{3} = \lambda_q
\,\epsilon^{ijk}\, \overline{u^c_L}_{i}\, d^{\phantom{c}}_{Rj}  S_k + \lambda_\chi
S^{*i}\bar\chi \,d^{\phantom{c}}_{Ri}
+\lambda_\phi \phi_a\bar\psi^a\chi+ {\rm h.c.} \ ,
\eea
where the complex scalar $\phi$ and the Dirac fermion $\psi$ are in the fundamental representation
of some strongly interacting  gauge group $G$. A confined bound state of $\phi$ and $\psi$ then
plays the role of the mirror nucleon in the model of Ref.~\cite{Berezhiani:2018udo}.\pagebreak

\noindent{\bf Nota bene} \\
In the ``Acknowledgments''  of
Ref.\cite{Berezhiani:2018udo}, the author states, in reference to his participation in the INT Workshop  ``Neutron-Antineutron Oscillations: Appearance,
Disappearance, and Baryogenesis'' \linebreak held October 23\,--\,27, 2017, that:\\
{\it However, my talk [43] evidently had some subconscious impact on
  the community, since very similar work of Fornal and Grinstein
  appeared recently [44], with the difference that the dark particle
  $n'$ was considered as an elementary fermion with ad hoc chosen mass,
  which was followed by many other works [45--55]. It is somewhat
  surprising that non of these authors mentioned about my talk, even
  those participants of the Seattle INT Workshop which were explicitly
  present on it -- something is rotten in the state of Denmark -- and
  such a solution for the neutron lifetime puzzle was coined as a the
  dark decay solution or Fornal-Grinstein solution, while it remains
  questionable whether it is a solution at all.} \\
\indent   We would like to take
this opportunity to clarify that neither Fornal nor Grinstein were
present at Berezhiani's third talk at the Workshop on October 26, 2017 (cited as [43] in Ref.~\cite{Berezhiani:2018udo}), the talk in which he
presented his ideas regarding neutron dark decay in the context of neutron-mirror neutron oscillations: 
Grinstein did not participate in the Workshop and Fornal had already left the Workshop on  October 24, 2017. 

We started working on our ideas  in May 2017, soon after the article by Geltenbort and Greene ``The neutron lifetime puzzle'' in the Institut Laue-Langevin 2016 Annual Report \cite{ILL},  based on Ref.\cite{Greene}, was brought to our attention. 
We have photographic evidence that our ideas leading to the publication \cite{Fornal:2018eol}
were already developed  by the time of the aforementioned INT Workshop. The blackboard photo \cite{photo} taken two weeks prior to the Workshop (the date can be checked in the file's properties) shows the diagram for the neutron dark decay that lies at the heart of our proposal. \\
\indent
One of us (BF) was first made aware of the content of
Berezhiani's third talk in an e-mail from Yuri Kamyshkov on January 20, 2018, i.e.~over two weeks after our paper \cite{Fornal:2018eol} was posted on the arXiv. 
Reference~\cite{Berezhiani:2018udo} seems to suggest that our work was somehow influenced by Berezhiani's talk. This is obviously not true. \\

\noindent{\bf Acknowledgment}\\
The authors were supported in part by the DOE Grant No.~${\rm DE}$-${\rm SC0009919}$.

\bibliographystyle{utcaps}
\bibliography{comm}

\end{document}